\documentclass[a4paper,twocolumn,spanish,english]{article}
\usepackage[T1]{fontenc}
\usepackage[latin9]{inputenc}
\usepackage{babel}
\addto\shorthandsspanish{\spanishdeactivate{~<>}}

\usepackage{textcomp}
\usepackage{url}
\usepackage{graphicx}
\usepackage[unicode=true,pdfusetitle,
 bookmarks=true,bookmarksnumbered=false,bookmarksopen=false,
 breaklinks=false,pdfborder={0 0 1},backref=false,colorlinks=false]
 {hyperref}
\usepackage{breakurl}

\makeatletter

\newcommand{\lyxaddress}[1]{
\par {\raggedright #1
\vspace{1.4em}
\noindent\par}
}

\AtBeginDocument{
  
}

\makeatother

\begin{document}

\title{A light tracking system to measure structural deformations}

\author{Damián Gulich,$^{1,2,3}$ \\
Mario Garavaglia,$^{1,2}$}

\maketitle

\lyxaddress{$^{1}$Centro de Investigaciones Ópticas (CONICET La Plata - CIC),
C.C. 3, 1897 Gonnet, Argentina}

\lyxaddress{$^{2}$Departamento de Física, Facultad de Ciencias Exactas, Universidad
Nacional de La Plata (UNLP), 1900 La Plata, Argentina }

\lyxaddress{$^{3}$Departamento de Ciencias Básicas, Facultad de Ingeniería,
Universidad Nacional de La Plata (UNLP), 1900 La Plata, Argentina}
\begin{abstract}
Light tracking systems have been used in recent years to study wandering
phenomena in atmospheric optics. We propose to employ this technology
in structural deformation sensing. 
\end{abstract}

\section{Introduction}

Light tracking systems have been used for years to study phenomena
in atmospheric optics \cite{ANDREWS,TESIS,ANGLE2006,ANGLE2007,DIFF2009}.
For example, behavior of spot wandering variance versus pupil diameter
in a laser propagation experiment is related to the structure constant
($C_{n}^{2}$) of the refractive index of air. In this situation it
is very important to have a high precision spot tracking system with
high sampling and recording frequency \cite{Rabal2010}. This technology
has been used in structural applications in the past \cite{CORTIZO}.
In this article we describe an application of this system to measure
sectional deformations in structures. Available electronics components
are detailed in Appendix \ref{sec:Opotoelectronic-system-for}.

\section{Basic setup}

\begin{figure}
\begin{centering}
\includegraphics[width=1\columnwidth]{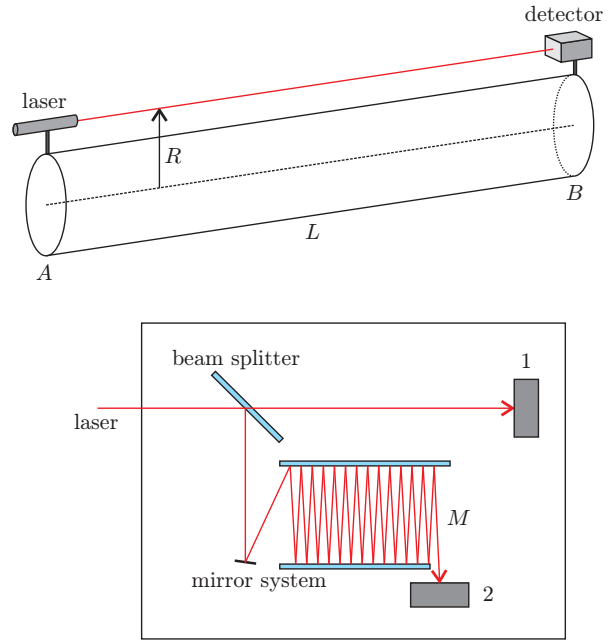}
\par\end{centering}

\caption{\label{fig:Basic-setup.}Basic setup and detector configuration. $M$
is the traveling distance of light from the beam splitter to light
tracker 2 after a system of mirrors (optical distance multiplier).}
\end{figure}

Suppose a long tubular of length $L$ and radius $R$ ($\sim$1.2
m in LIGO%
\footnote{Where temperature is constant and there is no atmospheric turbulence
of any kind since the whole system is kept in vacuum.%
}) structure suffering sectional types of deformation. Suppose section
$A$ is fixed and has a laser emitting with a proper focusing system
(see Fig. \ref{fig:Basic-setup.}). The registering station is fixed
at section $B$ (distance $L$) and both light trackers are centered. 

\subsection{$\varphi$-rotation}

\begin{figure}
\begin{centering}
\includegraphics[width=1\columnwidth]{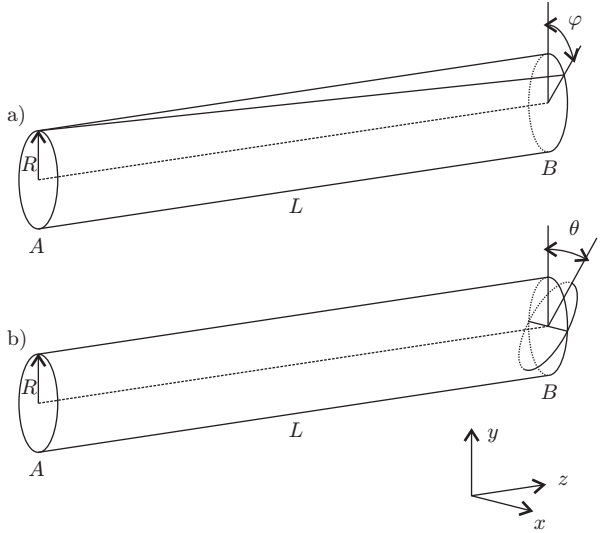}
\par\end{centering}

\caption{\label{fig:Rotations}Sectional rotational deformations.}
\end{figure}

If section $B$ rotates an angle $\Delta\varphi$ relative to section
$A$ (see Fig. \ref{fig:Rotations}-a), light tracker 1 will regard
it as a horizontal displacement $\Delta x$. Since $\Delta x\cong R\Delta\varphi$:
\begin{equation}
\Delta\varphi\cong\frac{\Delta x}{R}\label{eq:deltaphi1}
\end{equation}

In an ideal case, the smallest possible measurable displacement is
$\Delta x=2.54\cdot10^{-6}\mbox{ m}$. Let's assume that uncertainty
in $R=1.2\mbox{ m}$ is $\delta R=R/100$ ($\delta$ means the error).
Propagating errors (assuming $\Delta\varphi=0$) we find that 
\[
\delta\left(\Delta\varphi\right)\simeq\delta\left(\frac{\Delta x}{R}\right)\simeq2.117\cdot10^{-6}\mbox{ rad}
\]

Light tracker 2 will see it as an opposite and equal horizontal displacement
with a vertical component depending on traveling distance $M$.

\subsubsection{Comment: rotation around $y$ axis}

Suppose section $B$ rotates an angle $\Delta\psi$ around $y$ axis.
It will be seen by tracker 1 as a horizontal displacement and by tracker
2 as $M\times2\Delta\psi$ in opposite direction.

\subsection{$\theta$-rotation}

If section $B$ rotates an angle $\Delta\theta$ relative to section
$A$ (see Fig. \ref{fig:Rotations}-b), light tracker 2 will regard
it as a vertical displacement of magnitude $M\times2\Delta\theta$.
Light tracker 1 will see it as an opposite vertical displacement.

\subsection{Pure horizontal and vertical displacements}

For pure $y$ axis displacements both light trackers will detect a
vertical centroid displacement of the same magnitude. 

For pure $x$ axis displacements both light trackers will detect a
horizontal centroid displacement of the same magnitude but of opposite
sign due to beam splitter reflection in the registering station.

\section{Final remarks and analysis techniques}

Given the quality of this measuring setup it is possible to measure
an angular deviation from the laser station to the registering station
as small as $2.117\cdot10^{-9}\mbox{ rad}$ ($L=1\mbox{ km}$), as
well as other angular deviations in high frequency and precision.
Accuracy in other relative angular measurements can be greatly improved
by the optical distance multiplier ($M$ distance).

As for the analysis of time series, our group has experience with
wavelets \cite{ZUNINOwavelet}, information theory quantifiers (complexity-entropy
causality plane, etc.) \cite{ZUNINOcomplexity,ZUNINOinformation,ZUNINOmarkets}
and \emph{multifractal detrended fluctuation analysis} (MFDFA) \cite{GULICH}.

\rule[0.5ex]{1\columnwidth}{1pt}

\appendix

\section{Appendix: Opotoelectronic system for laser beam spot position tracking
\label{sec:Opotoelectronic-system-for}}

This system was developed and built at Laboratorio de Procesamiento
Láser (CIOp). It captures laser beam spot centroid positions at high
frequency \cite{Rabal2010} and records them to a PC. The system can
be operated with a single detector or with two in paralell. Sampling
rate for single-mode can be fixed up to 12000 samples/s and in dual-mode
it can be fixed up to 6000 samples/s for each detector.

\begin{figure}
\begin{centering}
\includegraphics[width=0.8\columnwidth]{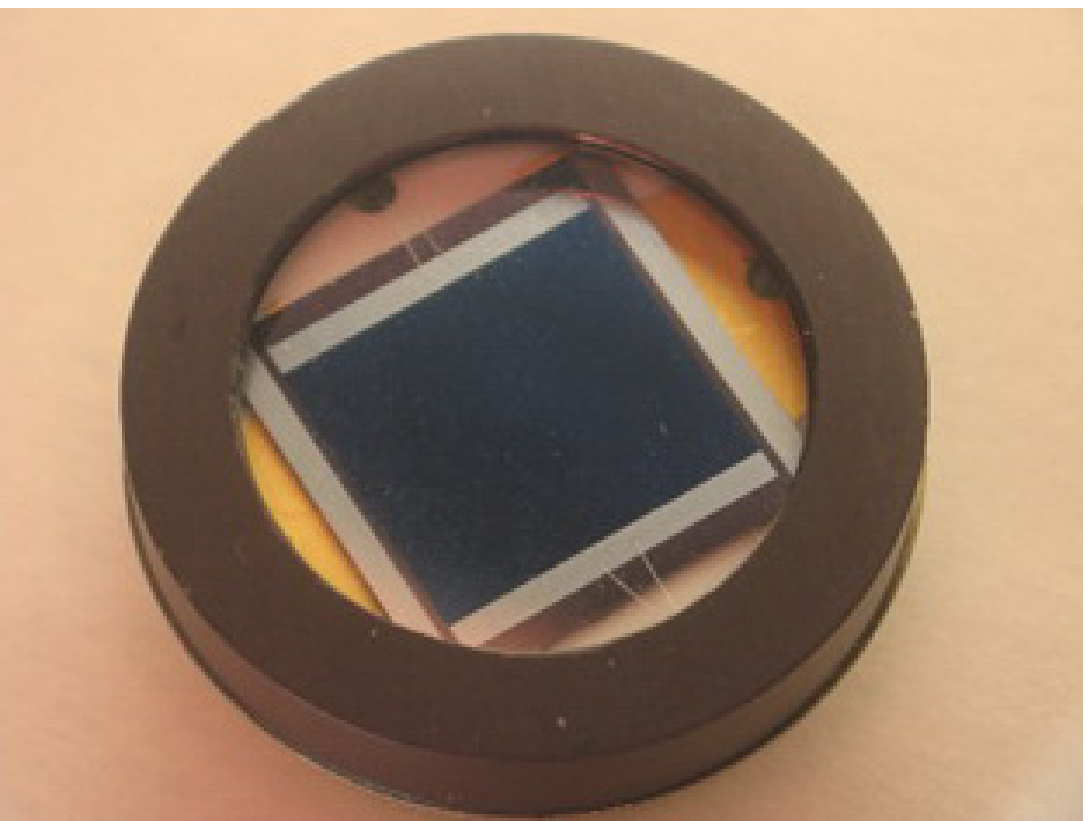}
\par\end{centering}

\caption{\selectlanguage{spanish}%
\label{fig:Detector-de-posicion}\foreignlanguage{english}{Continuous
position detector.}\selectlanguage{english}%
}
\end{figure}

\subsection{Technical specs for continuous position detectors}

\subsubsection{SC-10D \textendash{} Tetra-lateral}
\begin{itemize}
\item Responsivity \cite{TETRA}:

\begin{itemize}
\item min. 0.35 A/W.
\item typical 0.42 A/W.
\end{itemize}
\item Active area: 103 $\mbox{mm}^{2}$.
\item Dimensions: 10.16 mm $\times$ 10.16 mm.
\item Max. power density: 10 $\mbox{mW/cm}^{2}$.
\item Resolution: 0.00254 mm.
\item VBias: up to -15V.
\item Dark current:

\begin{itemize}
\item typical 0.025 $\mu\mbox{A}$.
\item max. 0.250 $\mu\mbox{A}$.
\end{itemize}
\end{itemize}
\begin{figure}
\begin{centering}
\includegraphics[width=1\columnwidth]{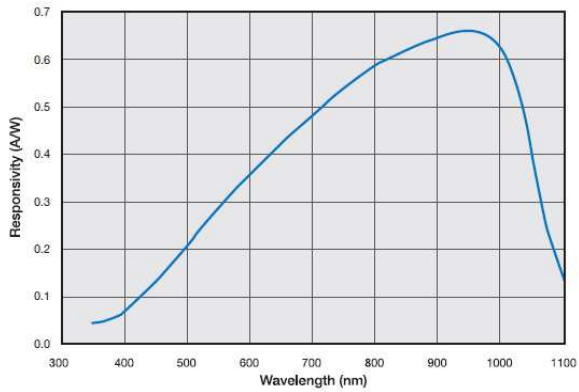}
\par\end{centering}

\caption{Typical spectral response for model SC-10D \cite{TETRA}.}
\end{figure}

\subsubsection{Model: DL-10 \textendash{} Duo-lateral}
\begin{itemize}
\item Responsivity \cite{DUO}:

\begin{itemize}
\item min. 0.3 A/W.
\item typical 0.4 A/W.
\end{itemize}
\item Active area: 100 $\mbox{mm}^{2}$.
\item Dimensions: 10 mm $\times$ 10 mm.
\item Máx. power density: 1 $\mbox{mW/cm}^{2}$.
\item Resolution: 0.00254 mm.
\item VBias: -5V.
\item Dark current:

\begin{itemize}
\item typical 500 nA.
\item max. 5000 nA.
\end{itemize}
\end{itemize}
\begin{figure}
\begin{centering}
\includegraphics[width=1\columnwidth]{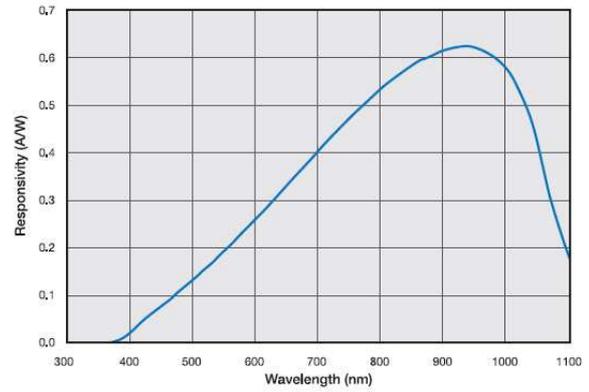}
\par\end{centering}

\caption{Typical spectral response for model DL-10 \cite{DUO}.}
\end{figure}

\subsubsection{Data acquiring board NI-USB6009}

Features \cite{NATIONAL}:
\begin{itemize}
\item 8 analog inputs, @ 14 bits and 48 kS/s. \emph{Differential mode} and
\emph{Single-ended}.
\item 2 analog outputs of 12 bits.
\item 12 digital input/output lines.
\item 32 bit counter, 5MHz.
\item Digital trigger.
\end{itemize}
\vfill{}

\end{document}